# On the analysis of the vibrational Boson peak and low-energy excitations in glasses

S. N. Yannopoulos[1,*], K. S. Andrikopoulos,[1,2] G. Ruocco[3,4]

[1] *Foundation for Research and Technology Hellas – Institute of Chemical Engineering and High Temperature Chemical Processes (FORTH / ICE-HT), P.O. Box 1414, GR–26504 Patras, Greece*

[2] *Physics Division, School of Technology, Aristotle University of Thessaloniki, GR–54124, Thessaloniki, Greece*

[3] *INFM-CRS SOFT, Università di Roma La Sapienza, 00185 Roma, Italy*

[4] *Dipartimento di Fisica, Università di Roma La Sapienza, 00185 Roma, Italy*

**Abstract**

Implications of reduction procedures applied to the low energy part of the vibrational density of states in glasses and supercooled liquids are considered by advancing a detailed comparison between the *excess* – over the Debye limit – vibrational density of states $g(\omega)$ and the frequency-reduced representation $g(\omega)/\omega^2$ usually referred to as the *Boson peak*. Analyzing representative experimental data from inelastic neutron and Raman scattering we show that reduction procedures distort to a great extent the otherwise symmetric excess density of states. The frequency of the maximum and the intensity of the excess experience dramatic changes; the former is reduced while the latter increases. The frequency and the intensity of the Boson peak are also sensitive to the distribution of the excess. In the light of the critical appraisal between the two forms of the density of states (i.e. the excess and the frequency-reduced one) we discuss changes of the Boson peak spectral features that are induced under the presence of external stimuli such as temperature (quenching rate, annealing), pressure, and irradiation. The majority of the Boson peak changes induced by the presence of those stimuli can be reasonably traced back to simple and expected modifications of the excess density of states and can be quite satisfactorily accounted for the Euclidean random matrix theory. Parallels to the heat capacity "Boson peak" are also briefly discussed.

1. **Introduction**

When considered from the viewpoint of vibrational dynamics, crystalline and non-crystalline (glasses and amorphous solids) materials exhibit noticeable differences at energies that are material dependent, but that always lie around 1 THz (1 THz ≙ 33.3 cm$^{-1}$ ≙ 4.13 meV ≙ 48 K). These differences, that have their origin at the lack of long–range periodicity, can all be traced back to the universal presence in glasses of low–energy excitations in excess of the parent crystals [1]. The term low–energy excitations has been coined to account for properties concerning: (i) the specific heat excess (bump in the $C(T)/T^3$ plot) at temperatures around 5-10 K, (ii) the plateau in thermal conductivity at about the same temperature, (iii) the strong attenuation of sound waves, and (iv) the presence of scattering intensity– absent from the parent crystal – in Raman and inelastic incoherent neutron scattering (INS) spectra [1-3].

The last issue, low energy scattering intensity excess, has recently resurfaced in view of new theoretical and experimental approaches [4-15]. In particular, crystalline solids obey rather accurately the predictions of the Debye model for an elastic continuum, according to which the vibrational density of states (VDoS) is a quadratic function of the frequency $g^D(\omega) \propto \omega^2$. On the contrary, non-crystalline phases ubiquitously exhibit some extra low-energy vibrational modes in excess to $g^D(\omega)$.

Historically, low energy excitations were first observed in Raman scattering as a low frequency peak in the spectrum of silica by Krishnan [16] who demarcated its boundaries in the interval 30-120 cm$^{-1}$. Few years later Flubacher *et al*. proposed the first interrelation between this low energy peak in Raman spectrum and the anomaly (non-Debye behavior) in heat capacity [17(a)]. The same anomaly was observed in heat capacity of crystalline silica. This was accounted for by invoking the idea of transversal mode flattening of the lowest transverse acoustic mode near the Brillouin-zone boundary, which results in a peak in the vibrational density of states (VDoS) [17(a)]. Interestingly, the same explanation was adopted for the low temperature anomaly in heat capacity of glassy silica due to an unusual flat dispersion [17(b), (c)]. Over the years, the terminology *Boson peak* was adopted in view of the fact that the temperature dependence of this low-frequency peak in the Raman spectra was found to follow rather well the Bose-Einstein statistics prediction concerning the thermal population of vibrational energy levels. Currently, the term Boson peak is, partly unduly, used for the peak in the $C(T)/T^3$ curve, owing to its possible relation with the vibrational modes that give rise to the excess scattering mechanisms.

Despite intensive studies by both experimentalists and theorists over the last three decades, the origin of the Boson peak in Raman and neutron scattering still remain a thorny issue of the non-crystalline solids physics. The inherent difficulty in the formulation of successful theories for glasses has led to the suggestion of numerous empirical correlations among different glass properties. Such correlations usually relate a spectral characteristic of the Boson peak (intensity, peak position etc.) with a specific property of the glass. Apart from the fact that most of those correlations are usually very scattered, devoid of predictive power, we will point out in the following sections that the use – of the often adopted – reduced VDoS can lead to erroneous conclusions.

An efficient way to obtain information on Boson peak is to study its spectral changes by changing the glass composition and/or the magnitude of certain external parameters. In the present paper particular attention will be paid to the comparison between the frequency-reduced VDoS and the true excess VDoS, i.e. the difference of the glass density of states and the Debye prediction. Then, based on the discussion of the various reduction procedures we will attempt to address some issues related to the dependence of the Boson peak intensity, position and shape under the influence of temperature, quenching rate, annealing, pressure and illumination in a variety of glasses.

## 2. Reduction schemes and their consequences in various aspects of the low-energy excitations phenomenology

The VDoS can indirectly be inferred in an inelastic (Raman) scattering experiment. Indeed, the spectral form of the first order vibrational part has been described – under certain approximations – by the following relation [18]:

$$I_{\alpha\beta}^{\exp t}(\omega) \propto C_{\alpha\beta}(\omega)g(\omega)\frac{n(\omega,T)+1}{\omega} \qquad (1)$$

where $I_{\alpha\beta}^{\exp t}(\omega)$ denotes the experimentally measured scattered Raman intensity; $C_{\alpha\beta}(\omega)$ is the photon–phonon or Raman coupling coefficient that is proportional to the scattering cross section of a vibrational mode at frequency $\omega$. The indices $\alpha\beta$ denote particular polarization geometries in a light scattering experiment. $n(\omega,T) = [\exp(\hbar\omega/k_B T) - 1]^{-1}$ is the Bose-Einstein occupation number with $\hbar$ and $k_B$ being the Planck and Boltzmann constants, respectively.

In an inelastic neutron scattering experiment one measures the dynamic structure factor $S(Q,\omega)$, where $Q$ is the momentum transfer and $\omega$ the energy transfer to the medium. Three main scattering mechanisms contribute to $S(Q,\omega)$, the elastic part, the one-phonon scattering mechanism, and the multiphonon background. In the incoherent approximation, the one-phonon term is related to the VDoS through:

$$S_{inc}^1(Q,\omega) \propto Q^2 e^{-2W} g(\omega) \frac{n(\omega,T)+1}{\omega} \qquad (2)$$

where $2W$ is the exponent of the Debye-Waller factor. It is worth noting that that Eqs. (1) and (2) are similar as regards the frequency dependence of the measured quantities, i.e. dynamic structure factor (neutrons) and first order vibrational intensity (Raman); the important difference is that Raman data are weighted by the coupling coefficient $C_{\alpha\beta}(\omega)$.

Equation (1) has formed the basis for the introduction of various spectral reduction approaches. As the different reduction schemes play a central role in the present paper, we will discuss them in some detail. The most frequently used are:

$$I_1^{red}(\omega) = \frac{\omega\, I^{expt}(\omega)}{[n(\omega,T)+1]} \propto g(\omega)\, C(\omega) \qquad (3a)$$

$$\chi''(\omega) = \frac{I^{expt}(\omega)}{[n(\omega,T)+1]} \propto \frac{g(\omega)\, C(\omega)}{\omega} \qquad (3b)$$

$$I_2^{red}(\omega) = \frac{I_{\alpha\beta}^{expt}(\omega)}{\omega\,[n(\omega,T)+1]} \propto \frac{g(\omega)\, C(\omega)}{\omega^2}. \qquad (3c)$$

$I_1^{red}(\omega)$ is the reduced Raman spectrum which isolates the VDoS weighted by the coupling coefficient. It singles out the trivial thermal population factor effects in the low frequency part. The reduced Raman intensity given by Eq. 3(b), $\chi''(\omega) = I_1^{red}(\omega)/\omega$, is the susceptibility representation and forms the common basis for analyzing relaxational effects revealed in depolarized light scattering spectra especially in the course of mode coupling theory [19]. Finally, Eq. 3(c) has been introduced in order to restore the experimentally measured well-resolved Boson peak in Raman scattering. In practice, $I_2^{red}(\omega) \propto T^{-1} I_{\alpha\beta}^{expt}(\omega)$ since $[n(\omega,T)+1]^{-1} \approx \hbar\omega/k_B T$ for $\hbar\omega/k_B T << 1$. In essence, transformation 3(c) leaves the frequency dependence of the experimentally measured spectrum almost intact. An illustrative paradigm of the various reduction schemes are shown in Fig. 1 applied to the room temperature depolarized Raman spectrum of vitreous silica. This comparison reveals that the VDoS (modulated by $C(\omega)$) obtained by the genuine reduced representation [Eq. 3(a)] is featureless in the region (indicated by the arrow) where the experimental data show the Boson peak. This implies that the true excess (over the Debye level) $g(\omega) - g^D(\omega)$ (*excess VDoS* or *e-VDoS*) of vibrational modes in the low energy spectral region is indeed small. The same trend is exhibited in the VDoS as obtained by INS. Obviously, the intensity and the energy of the Boson peak which appears in the experimental data as well in the $I_2^{red}(\omega)$ representation is influenced by the presence of the $[n(\omega,T)+1]/\omega$ factor and the $\omega^2$ term, respectively. This fact had already been noticed since the first studies of vibrational states in amorphous solids by Brodsky *et al*. [20] who stated that *the low frequency peak is an artifact of the frequency and temperature factors*.

To retain the Boson peak, the reduced representation Eq. 3(c) is usually adopted in the analysis of Raman data and analogously the frequency-reduced VDoS (fr-VDoS) i.e. $g(\omega)/\omega^2$ or $g(\omega)/g^D(\omega)$ in INS. The former, $g(\omega)/\omega^2$, is usually preferred since the calculation of the Debye VDoS requires either the knowledge of the VDoS of the corresponding crystal or can be in principle estimated in its low frequency region from longitudinal and transversal sound velocity data at the same temperature. This has also stand as another obstacle in determining the e-VDoS $g(\omega) - g^D(\omega)$ which contains the accurate information concerning the excess modes.

## 3.     Frequency-reduced VDOS vs. excess VDOS: experimental facts and theoretical predictions

The dogma of the low-energy phenomenology on amorphous solids is that non-crystalline phases exhibit *prominent* differences in their low energy region (1-5 meV) in comparison to the corresponding crystalline phases or the Debye spectrum. However, a comparison of the VDoS in the full energy range [see for example the case for otrhto-terphenyl (oTP) (Fig. 2 in [21]) and toluene (Fig. 2 in [22])] reveals that the difference $g(\omega) - g^D(\omega)$ in this region is negligible compared to the modifications that VDoS experiences at higher frequencies. Actually, the VDoS of the glass and the crystal become distinctively different if one attempts the comparison in the fr-VDoS representation, i.e. in the function $g(\omega)/\omega^2$. However, the question that arises is: which of the two aforementioned combinations of the vibrational density of states, i.e. e-VDoS or fr-VDoS, is fundamentally important?

Undoubtedly, the important quantity that is genuinely related to the properties of a glass is the VDoS itself or the excess VDoS, i.e. $g(\omega) - g^D(\omega)$ because this quantity represents the number of vibrational modes at a specific frequency interval (given in units of energy$^{-1}$). On the other hand, the fr-VDoS quantifies the ratio of the excess vibrational states at each frequency over the corresponding Debye expectation. In this sense fr-VDoS (given in units of energy$^{-3}$) is a kind of an *amplitude histogram*. Despite the convenience provided by the fr-VDoS to easily reveal the Boson peak, this function conceals a large piece of information conveyed by e-VDoS. Obviously, the use of the $g(\omega)/\omega^2$ representation is justifiable when one needs to emphasize the deviation of the low energy glass VDoS from the corresponding Debye trend, but can lead to misleading conclusions if the Boson peak positions and intensities in fr-VDoS's are used in a comparative manner. This is due to the division by $\omega^2$ which unavoidably (and artificially) enhances and shifts towards lower frequencies the peak that corresponds to the e-VDoS, see Fig. 5 in Ref. [9]. A characteristic problem caused by the division with the $\omega^2$ term is that Boson peaks seem to exist also in crystals, e.g. in shape memory alloys [23], in zeolites [14, 24], as well in several other cases of materials in the crystalline state especially in polymers. Obviously, this adds some confusion to the interpretation of the Boson peak and its relation to glass properties. Below we present some representative examples of the inconsistencies caused by the use of fr-VDoS.

Figure 2 shows an example of the various forms of the VDoS for the elemental glass *Se*. The density of states for the glass and the crystal are provided by INS [25]; the upper panel shows the excess and the fr-VDoS. This figure reveals two important features about the e-VDoS: (i) it is a rather symmetric curve in contrast to the much skewed form of the fr-VDoS, and (ii) its maximum is situated at a much higher frequency than that of the fr-VDoS. To demonstrate that the above mentioned observations as regards *g-Se* are not glass specific but are rather general we illustrate in Fig. 3 the e-VDoS estimated for various glasses, (a) $B_2O_3$ [26], (b) oTP [21], (c) polybutadiene [27], and (d) glycerol [28] by subtracting the Debye contribution from the experimentally determined VDoS. Again the analysis shows that the e-VDoS is a rather symmetric peak located at higher energy than that of the peak in the fr-VDoS by a factor of two. Better inspection of the data reveals a slight asymmetry, i.e. the low frequency side of the peak seems slightly more extended than the high frequency one. In the case of polybutadiene and glycerol, data for various temperatures are available. It is interesting to notice that the e-VDoS of PB exhibits a *decrease* in intensity and a *red-shift* in energy *with increasing temperature*. The latter is not surprising since temperature rise is expected to weaken the "vibrations" associated with the excess modes. This behavior is reminiscent of the predictions of the Euclidean random matrix theory (ERMT) [6] where the spectral density of the excess modes shifts systematically to lower energies as temperature increases. On the contrary, the decrease in intensity is a non-trivial outcome. At a first glance it seems surprising since with increasing temperature one expects a growing intensify of the excess modes. However, the results demonstrate that the contribution of the Debye spectrum, grows with a faster rate, and thus can account for the grater part of the VDoS increase. This is an important observation that usually goes unnoticed. An even more important consequence of this observation is that the impression of a vividly growing quasi-elastic line above the glass transition temperature becomes highly questionable. At this point one could claim that the decrease in the excess, Fig. 3(c), is due to the migration of a fraction of the excess modes to negative part of the spectrum of the eigenvalues (unstable modes); however, such behavior is expected to occur above some critical temperature (associated with the $T_c$ of mode coupling theory in the context of the ERMT) usually located within the supercooled regime. This does not apply for the cases under study since the decrease of the intensity of the excess modes starts at much lower temperature, essentially within the glassy state. The same effect has also been revealed in detail in studies of the low energy modes of the vitreous, supercooled and molten silica [29]. The results for silica is even more intriguing since the energy of the Boson peak itself (i.e. the peak in the fr-VDoS) anomalously increases with temperature rise while the energy of the peak representing the spectrum of the excess modes was found to follow the opposite trend.

The experimental facts presented above unambiguously demonstrate that the Boson peak observed in the fr-VDoS has no common features with the true excess VDoS. The reduction by the $\omega^2$ term seriously distorts the rather symmetric envelope of the excess modes thus leading to the putative "universal" form of the Boson peak. Moreover, the actual energy of the maximum of the e-VDoS peak is always higher than the Boson peak energy by more than a factor of two. This raises doubts on the estimations of structural correlations lengths characteristic of the length scale of the medium range structural order, which are based on the Boson peak frequency. Forced by the above observations we proceed below to some quantitative comparisons between the two forms of the density of states, i.e. the e-VDoS and the fr-VDoS, which will make clear the problems originating from the use of the latter in the comparison of Boson peaks as a function of some external parameter or between different glasses.

Figure 4 reports a sketch, where different schematic models of excess of states are added to a "normal" Debye VDoS and represented in the e-VDoS and the fr-VDoS format. In panel (a) we show different curves where a certain (constant area) fraction of excess modes, distributed following a Gaussian shape, has been superimposed on a Debye background; only the position of the peak maximum of the various e-VDoS's has been systematically shifted. This effect somehow accounts for the mode softening observed with increasing temperature in a real glass. In reality the Debye spectrum also changes, but for simplicity we keep it constant here since it will not influence the effect we attempt to demonstrate. In panel (b) we show the frequency-reduced form of the composite VDoS curves mentioned above. Four major observations can be made out from the obtained fr-VDoS curves. (i) The peak maxima of corresponding curves in e-VDoS and fr-VDoS differ appreciably with the latter being located at much lower frequencies. (ii) Although the e-VDoS was chosen to be symmetric, fr-VDoS exhibits a remarkable degree of asymmetry. (iii) The integrated area of the peaks in the fr-VDoS representation increase in a systematic way as the peak maximum of the e-VDoS is shifted to lower frequencies, despite the fact that the integrated areas of the e-VDoS peaks were set equal. (iv)

The limiting energy $\to 0$ value of the fr-VDoS exhibits an increasing trend, reminiscent of the quasi-elastic scattering component in the low frequency Raman spectra, which experimentally is found to grow rapidly when heating the glass above the glass transition temperature. To check if these results are dependent upon the particular form (Gaussian, concave-like tails) of the e-VDoS peak we present in Fig. 4, panels (c) and (d), another example where the e-VDoS has a different shape with convex-like tails. The corresponding fr-VDoS curves shown in (d) also exhibit the four aforementioned features. Interestingly, in this case the peaks in fr-VDoS bear a remarkable resemblance to the experimentally measured Boson peaks. The peaks in (d) are much more asymmetric and the peak maxima between e-VDoS (c) and fr-VDoS (d) differ by a larger amount that the corresponding ones in (a) and (b). A comparison between the two cases, concave and convex tails of the e-VDoS reveals that the above observations are not dependent on the specific form of the chosen e-VDoS.

Comparing the experimental data presented in Figs. 2 and 3 and the results of the model curves given in Fig. 4 one can easily discern a similarity between the experimental e-VDoS in structural glasses and the e-VDoS of the model curves in the convex-like example. This observation is supported by recent theoretical approaches that predict the energy dependence of the excess spectrum. Indeed, in the framework of the Euclidian random matrices approach Parisi and coworkers [6] studied the inherent structures (local minima of the potential energy) where the Boson peak was connected with a phase transition from a minima-dominated phase (with phonons) at low energy to a saddle-point-dominated phase (without phonons). According to the analytical result of the ERMT, the spectrum of the harmonic oscillations around (i) the inherent, (ii) the generalized inherent structures, and (iii) the instantaneous configuration is of a semicircular shape whose edges are a function of the temperature. Fig. 5 depicts a schematic representation of those predictions. Actually, the e-VDoS possesses the semicircular – and hence totally symmetric – form when plotted against the vibrational *eigenvalue* $\lambda \propto \omega^2$ (see lower inset in Fig. 5). It is interesting to note that the e-VDoS spectrum vs. $\omega$ exhibits a slight asymmetry on the low energy side of the peak; this observation is in line with the experimental observations as mentioned above (see the e-VDoS) in Figs. 2 and 3.

Closing this paragraph, it would be instructive to note that a Gaussian-like (concave tails) excess of modes was found in the experimental VDoS case of doped crystals [30] determined by INS, which exhibit some excess density of states at low frequencies. A localized Einstein mode in addition to the Debye spectrum of the crystal was found to describe the experimental data quite satisfactorily. Obviously, calculating the fr-VDoS of these doped crystals a Boson peak-like feature will appear in a purely crystalline solid. This is a further indication of the problems that can arise from the use of the frequency reduction of the vibrational density of states.

**4. Temperature-induced spectral changes of Boson peak**

*4.1 The temperature dependence of the e-VDoS*

Temperature is a parameter that is frequently used as a control variable in studies of the low energy excitations in solids. The generally observed temperature induced changes in the frequency-reduced ($I_2^{red}$) low energy spectra are manifested as an almost temperature insensitive part of the spectrum above the Boson peak energy and a systematically growing intensity below this energy. An exception to this is vitreous silica where the above behavior is reversed; the reduced spectra show a decreasing intensity with temperature rise at least up to the melting point [29]. The general picture mentioned above, is usually considered as evidence for the harmonic nature of the Boson peak and the anharmonic nature of the quasi-elastic line contributing in the very low frequency region of the spectrum. Taking into account the discussion advanced in the previous section, and in particular the fact that the Debye level is a function of temperature, the consideration of a strong rise of the quasi-elastic line is not necessary. Indeed, the Debye contribution is rapidly growing, especially above the glass transition temperature, and hence can account for the increasing intensity of the low frequency spectrum, see for example Fig. 4 in [31].

As regards the temperature dependence of the e-VDoS, see Fig. 3(c) for polybutadiene and (d) for glycerol, the experimental data show that this quantity is a rapidly decreasing function of temperature. At the highest temperature for both glasses the reduced density of states $g(\omega)/\omega^2$ exhibits a drastic increase as $\omega \to 0$, i.e. the quasi-elastic line. In the e-VDoS representation, apart from the intensity reduction of the excess modes there is also a shift towards zero. This continuous shift will lead above a certain temperature, characteristic of the material, to a non-zero intercept at $\omega = 0$ signifying the instability of the involved modes. In this case, the division by $\omega^2$ will give rise to a strong intensity in the frequency range below the energy of the Boson peak. ERMT predicts the shift of the e-VDoS to lower energies. More specifically, the calculations for the lowest energy eigenvalue, i.e. the show linear temperature dependence [6]. Using the limited data of Fig. 3 we present in Fig. 6(a) the temperature dependence of the eigenvalue of the excess modes for polybutadiene and glycerol. Since the lowest eigenvalue in the experimental data is not well resolved, especially at high temperatures, we have chosen to present the eigenvalue at the maximum of the e-VDoS which is also expected to follow the temperature dependence of the lowest one. The solid and dashed lines are linear fits to the experimental data.

*4.2   Quenching rate dependence and annealing effects on Boson peak*

Annealing effects on the spectral feature of the Boson peak can also be accounted for by reflecting the spectral changes to the e-VDoS. Early observations of the temperature dependence of the frequency of the Boson peak in Raman scattering showed that the faster the quenching of $As_2S_3$ glass the lower the frequency of the Boson peak [32]; further, annealing of slowly quenched glasses resulted in a further increase of the Boson peak energy. Annealing effects on polymeric glasses, i.e. poly(methyl methacrylate) (PMMA) studied by INS [33] and Raman scattering [34] showed that the Boson peak intensity decreases while its position increases after annealing near the glass transition temperature. Estimating the VDoS from heat capacity data, in $B_2O_3$ glass, it was found that the magnitude of the lower-energy side of the Boson peak decreases and the energy of the peak maximum increases with the progression of annealing [35]. In contrast to the above studies, an INS study of the low energy excitations in polybutadiene [36] revealed no detectable changes in the VDoS of the this glass after annealing. Summarizing, in all cases where annealing was found to induce some changes of the Boson peak these changes were located in its low energy side and in essence involved a blue-shift of its frequency and a decrease of its intensity.

The above short survey on some experimental findings of annealing effects shows that the situation is not yet clear. In the case of Raman studies one should know the way the coupling coefficient is affected by the thermal history (annealing or quenching). More importantly, even if we accept that the changes in intensity and energy of the Boson peak are real, it cannot be unambiguously claimed that the intensity of the Boson peak decreases. The reason is that the Debye VDoS also changes after annealing/quenching and hence the net effect on the excess VDoS is not known. Actually, knowing that the sound velocity increases with annealing it is obvious that the Debye contribution to the total VDoS will decrease after annealing. Hence, the decrease of the total VDoS observed experimentally does not necessarily imply a decrease of the modes (e-VDoS) responsible for the Boson peak. It is therefore suggested that a comparison of the total VDoS and the Debye level should be attempted before concluding about the modifications of the vibrational modes responsible for the Boson peak. However, what can be rather safely stated is that annealing induces stabilization of some unstable modes trapped in metastable configurations in the course of fast quenching. Consequently, a certain fraction of the spectrum of the excess modes will migrate to higher frequencies, in accordance to the ERMT, and hence the reduction by the $\omega^2$ term will result in a decreased intensity if the fr-VDoS.

*4.3. Hyperquenched glasses*

It has been recently observed by Angell and co-workers [37] that the Boson peak intensity in a hyperquenched (HQ) glass is more intense than that in the normal or annealed glass. The explanation of this observation is that the rapid quenching leads to a low temperature glass which retains the structural features and hence the details of the potential energy surface (PES) pertaining to the high temperature liquid state: the highest the quenching rate, the highest the energy in the PES where the system is trapped and hence the highest the "effective" temperature. However, as has been repeatedly mentioned in the foregoing discussion, in order to conclude about the intensity change of the Boson peak, one should compare the e-VDoS of the HQ and annealed glass. This is quite important in such studies since the sound velocity of the HQ glass (liquid-like structure, softer elastic constants) is expected lower than that of the slowly quenched or annealed glass. As a result the Debye contribution will inevitably be higher in the HQ and hence one cannot safely conclude about a net increase of the total VDoS.

Hyperquenching effects on the Boson peak of vitreous silica would be an interesting study in view of the anomalous temperature dependence of both its intensity and position [29]. Preliminary Raman studies on HQ silica fibers [38] have shown that annealing has the effect to decrease the frequency, and at the same time to increase the intensity of the Boson peak, tending to bring both to the bulk silica glass values. This observation is in the opposite way with the changes induced by annealing in other glasses (see section 4.2). Obviously, this behavior has its origin at the anomalous temperature dependence of the low energy excitations in vitreous silica. More systematic studies are in progress to clarify the hyperquenched-induced changes in this glass.

*4.4. Resolvability of the Boson peak in the molten state*

The fact that for several materials the Boson peak is well resolved, i.e. the minimum at its low energy side is preserved, even when the glass-forming material is driven to the equilibrium liquid state [29] is an important piece of information that can be used to elucidate its structural origin. This observation casts serious doubts on the consideration of a vividly increasing quasi-elastic line with temperature rise and on the medium-range structural order origin of the Boson peak. Bearing in mind that those glasses for which the Boson peak survives in the melt are network-forming glasses and considering the predictions of the ERMT [6] the resolvability of the Boson peak in the liquid state can be reasonably rationalized, see [29] for details. Further, it is worth to refer to a recent glass transition theory elaborated by Tanaka, i.e. the two–order–parameter model of liquids [39]. In this context, two types of structures are favored in liquids; normal liquid structures governed by the density order parameter and locally favored structures governed by the bond order parameter. Among other issues the two–order–parameter model has also been used to rationalize the experimental findings of Boson peak existence in the normal liquid state [39]. The physical origin of the Boson peak was ascribed to cooperative vibrational modes localized resonantly to locally favored structures that possibly exist above the melting point.

**5.     Pressure–induced spectral changes of Boson peak**

Pressure is another important external parameter that can appreciably affect the vibrational dynamics in glasses. It provides the advantage to single out the effect of density since in the case of temperature variation both density and thermal effects are involved. On the experimental side, the studies of pressure-induced changes on Boson peak are sporadic in contrast to studies of temperature-induced structural changes due to inherent experimental difficulties. In particular, in Raman scattering the application of very high pressures (up to tens of GPa) is relatively trivial; however, in this case the windows of pressure cells cause problems in the resolution of the scattered light near the elastic peak and further make not possible polarization studies. In INS, the need for bulky (massive) samples poses limits to the applied pressure (of the order of 1 GPa). The wide range of pressure application which is possible in Raman scattering is essential for checking the predictive power of theories dealing with pressure effects on Boson peak [40].

Pressure studies conducted so far, both experimentallly and with the aid of simulations (see the relevant references in [2] and [40]), show, with increasing pressure, an increase of the Boson peak energy and a concomitant reduction of its intensity. This is usually accounted for by considering a suppression of the modes responsible for the Boson peak. However, as in the case of temperature-induced changes discussed above, the influence of pressure on the Debye VDoS is completely overlooked. It is interesting to note that $g^D(\omega) = 3\omega^2 / \omega_D^3$ diminishes rapidly with the increase of the Debye frequency $\omega_D^3 \propto \rho v_\ell^3 v_t^3 /(2v_\ell^3 + v_t^3)$ where $\rho$ is the density. It is therefore not clear if the suppression of the VDoS of the glass under pressure comes from a real suppression of the excess modes responsible of the Boson peak or from the drastic decrease of the Debye level. In addition to this, we should bear in mind that a possible shift of the e-VDoS spectrum to higher $\omega$ under the application of pressure has also as a result the reduction of the Boson peak intensity which is proportional to $g(\omega)/\omega^2$.

Let us attempt an estimation of $g^D(\omega)$ reduction under pressure for GeS$_2$ for which both Raman data and sound velocities are available [41]. From Raman spectra we estimated the $g(\omega) \times C(\omega)$ product at a certain frequency (we chose the zero-pressure Boson peak frequency) which as a function of pressure is shown in Fig. 7 normalized to unity at $P = 0$. The Debye level was calculated from the pressure dependence of the sound velocity assuming a similar dependence of both the longitudinal and the transversal. It is obvious from Fig. 7 that the suppression of $g^D(\omega)$ is stronger compared to the suppression of the total VDoS. Considering the increase of density under compression it turns out that the decrease of $g^D(\omega)$ would be even stronger. Unfortunately, we have no estimation of the pressure effect on the Raman coupling coefficient, which however is not expected to change so remarkably. It is worth to remind, however, that the coupling coefficient is proportional to the derivative of the atomic/molecular polarizability with respect to the interatomic/intermolecular distance and/or relative orientation. This derivative is thus particularly sensitive to the local structure, being, for example, vanishing in the limit of a symmetric local environment. Therefore, we expect a suppression of its magnitude with pressure application. If true, this means that the reduction in $g(\omega)$ would be even smaller than the reduction in $g^D(\omega)$. Therefore, these data show that pressure application has dramatic effects on the Debye contribution while the excess VDoS should *intensify* with pressure increase in order to account for the milder drop of the total VDoS in comparison to $g^D(\omega)$ in GeS$_2$. Similar results are obtained for As$_2$S$_3$ glass after a comparison of the total and the Debye density of states [42]. On the contrary, the opposite effect is observed in densified vitreous silica. The total VDoS does indeed decrease with increasing density [43]. A change in density from 2.20 to 2.63 g cm$^{-3}$ correspond to a pressure change form 0 to 6 GPa [44]. The sound velocity in vitreous silica anomalously drops with increasing pressure exhibiting a minimum at ~3-4 GPa; at 6 GPa is still lower than the zero pressure value and hence $g^D(\omega)$ increases. Therefore, the decrease of the total VDoS and the increase of $g^D(\omega)$ imply a net suppression of the e-VDoS in silica. Concluding, in order to clarify the influence of pressure on the Boson peak or the excess VDoS the pressure-induced changes of $g^D(\omega)$ must also be taken into account.

### 6. Light illumination-induced effects on Boson peak: Athermal structure softening

Illumination-induced changes in structural, optical, rheological, mechanical, etc. properties of amorphous materials is a well established field of research. The majority of studies concern chalcogenides glasses which are amorphous semiconductors [45]. A remarkable photo-induced effect pertains to the athermal photo-induced fluidity (PiF) [46]. In the PiF regime a glass can "flow" under the combined action of sub-bandgap light illumination and mechanical stress, without heat provision. The effect takes place at temperatures much below the glass transition temperature [47]. PiF gave evidence that glass transition dynamics in glasses can be induced athermally by sub-bandgap light illumination. Its athermal nature became evident from the anomalous temperature dependence of the PiF, namely the fact that the photo-induced flow becomes hindered at higher temperatures [48]. The Boson peak changes of As$_2$S$_3$ in the course of PiF have already been noticed; see Fig. 5 in Ref. [47]. In brief, that study revealed that the frequency of the Boson peak experiences no changes as a function of stress while its intensity was found to alter according to the data presented in Fig. 8 which shows the ratio of the Boson peak intensity to the intensity of the main peak and this ratio is normalized to unity for the bulk, not-illuminated glass. Since there is no shift of the Boson peak we might conclude that the frequency of the e-VDoS from which it originates should also be stress-insensitive. Therefore, intensity changes of the Boson peak can be reflected to intensity changes of the e-VDoS. Of course the role of the coupling coefficient is unknown; under the assumption that it does not appreciably changes we can rationalize this effect bearing in mind that chalcogenide glasses, apart from covalent bonds, responsible for the glass structure robustness, possess "soft" or "defect" bonds that are easily amenable to external stimuli, such as light. Further, photo-plastic effects (i.e. PiF) involve two steps: (a) soft bonds are athermally broken under sub-bandgap light illumination, and (b) the applied stress exploits the soft bond rupture inducing structure deformation. Therefore, the reduction of defect concentration during PiF leads to a stiffer or more "ordered" structure with fewer soft or excess modes. This observation and the explanation given above, is in line with a recent study [49] where some similarities of the Boson peak in athermal bond breaking by changing the glass composition and temperature-induced softening in constant composition was proposed.

Illumination effects have also been studied in $As_2S_3$ glass [50] as a function of time by constant irradiation with sub-bandgap light. The effect on the Boson peak spectral features was analogous to that produced by thermal annealing, i.e. shift of the peak to higher energies and reduction of its intensity. Germania-doped silica glasses (5% and 13% $GeO_2$) have also been investigated under UV illumination [51]. For both samples, the Boson peak exhibited a shift to higher energies while its intensity showed opposing behavior, i.e. it increased for the low-doped whilst it decreased for the highly-doped glass. Again, the change of the coupling coefficient is not known and hence illumination induced studies cannot be definitely conclusive.

### 7. "Boson peak" and the low temperature heat capacity

The conclusions of the critical assessment presented in this paper about the Boson peak could probably be valid also for the "Boson peak" in heat capacity data, which is manifested in the $C(T)/T^3$ function. In this case the reduction by the T$^3$ term induces even more severe distortions to the rather symmetric excess $C(T) - C^D(T)$. Figure 9 illustrates two representative examples of heat capacity data of a polymer [52] and an inorganic glass [53] analyzed in the spirit of the ideas developed in this paper. The outcome concerning the true excess in heat capacity is quite similar to that found for the e-VDoS: the asymmetry comes from the T$^3$ term and the area under the $C(T)/T^3$ peak depends on the exact position of the excess heat capacity. In a nutshell, the use of $C(T)/T^3$ data in comparison the "Boson peaks" of various glasses could be misleading.

### 8. Conclusions

The existence of the Boson peak, i.e. the low frequency peak in the frequency-reduced VDoS $g(\omega)/\omega^2$, creates a conundrum for the physics of glasses. How it behaves under external stimuli and, more importantly, what is its origin are still unsettled questions. In this paper we attempted to face these issues from a different viewpoint than that usually followed. For (some of) those glasses for which the frequency-reduced VDoS is known, we derived the true excess of VDoS over the Debye level. In all cases, the latter quantity turns out to be a rather symmetric band, peaked at much higher energy than that of the Boson peak. The critical comparison of the true excess with the frequency-reduced density of vibrational states indicates that the use of the second representation does indeed distort the spectral features of the excess to a dramatic extent. The following conclusions can be drawn from the observations reported in this paper.

(i) A symmetric distribution of excess states gives rise to an asymmetric Boson peak. The observed (universal) asymmetry of the Boson peak would therefore be an artifact, and this would lead to conclude (at least based on Boson peak data) the inconsistency of the hypothesis that clusters with asymmetric size distribution exist in glasses.

(ii) The position of the maximum of the Boson peak is not directly related to the maximum of the distribution of the excess of states. On the contrary, it turns out to be sensitive to the type of the tails of the low frequency edge of such a distribution. It is therefore meaningless to use the Boson peak frequency to estimate the correlation length characterizing the medium range order of the glasses.

(iii) Various conclusions have been reached in the past on the glass microstructure based on the dependence of the Boson peak intensity on external stimuli (temperature, quenching rate, annealing condition, thermal history, pressure, irradiation, etc.). However, since the Boson peak intensity changes are not only related to the change of the number of excess modes, but also to their frequency location and to the details of their distributions, these conclusions must be considered with care.

Finally, the predictions of the Euclidean random matrix theory [6] seem to be a very satisfactory rationale for the experimental facts concerning the excess of states (in particular its shape and its temperature dependence) as well as the visibility of the Boson peak in the normal liquid state. The arguments developed for the VDoS seem to have their analogue for the "Boson peak" of heat capacity.

**Figure Captions**

**Fig. 1:** Comparison of the various reduction schemes given by Eqs. 3(a, b, c) applied in the depolarized low frequency Raman spectrum of vitreous silica. The intensities of the spectra have been normalized so as to facilitate the comparison.

**Fig. 2:** Lower panel: Low energy VDoS of g-Se and c-Se. Upper panel: excess and frequency-reduced VDoS of g-Se.

**Fig. 3:** Excess VDoS for four glass-sforming substances. In (a) and (b) the fr-VDoS has also been included for to comparison.

**Fig. 4:** Sketch of different schematic models of excess of states superimposed on to a "normal" Debye density of states. See text for details.

**Fig. 5:** Theoretically predicted spectrum of the excess VDoS by the ERMT. The upper insets show the frequency-reduced form of the e-VDoS.

**Fig. 6:** Temperature dependence of the eigenvalue of the maximum of the e-VDoS for glycerol and polybutadiene. The lines represent linear fits according to the ERMT.

**Fig. 7:** Pressure dependence of the total VDoS (weighted by the coupling coefficient) and of the Debye VDoS for $GeS_2$ glass. The more rapid drop of the latter does not provide supportive evidence for the suppression of the modes responsible for the Boson peak.

**Fig. 8:** Changes of the Boson peak intensity in $As_2S_3$ glass in the course of sub-bandgap light illumination and mechanical stress. See text for details.

**Fig. 9:** A comparison of heat capacity data in reduced ("Boson peak") and excess representation.

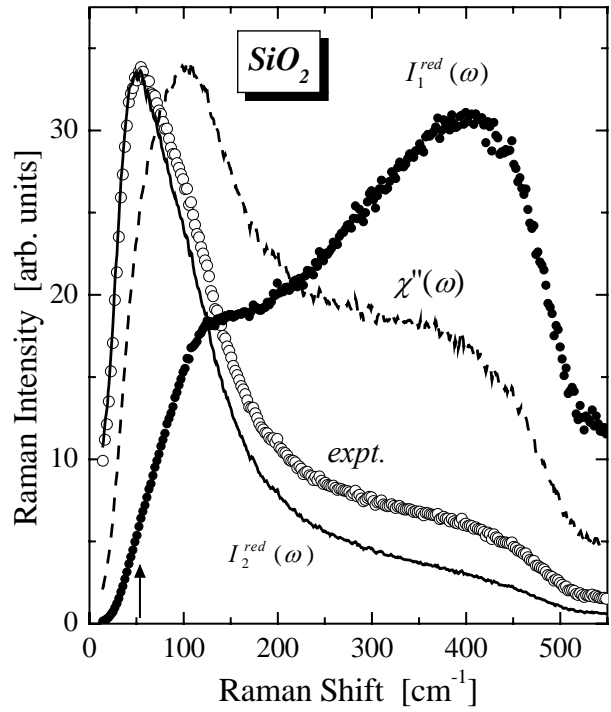

**Fig. 1**

*Reduction schemes of Boson peaks…*

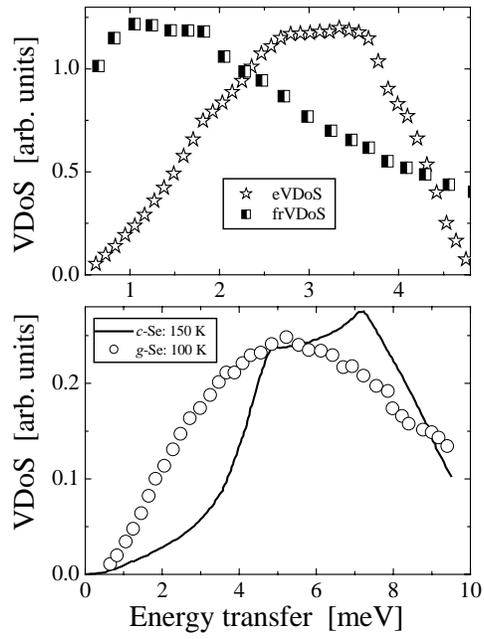

**Fig. 2**

*Reduction schemes of Boson peaks…*

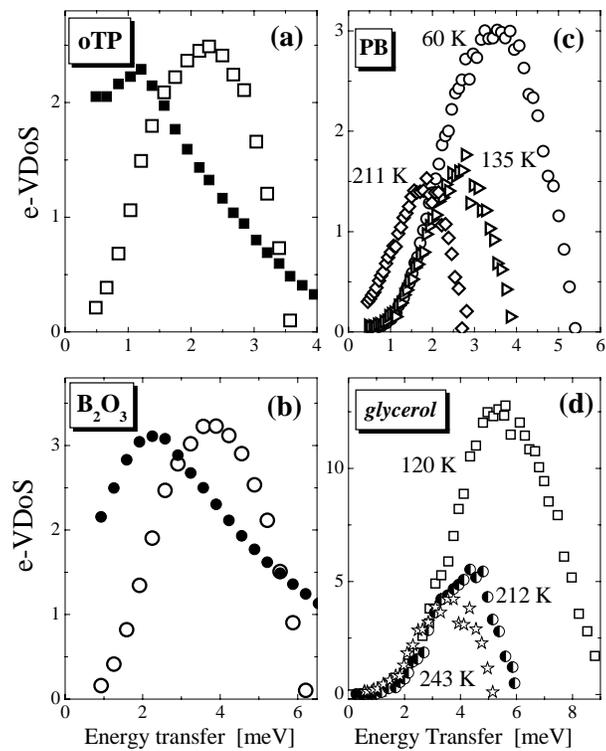

**Fig. 3**

*Reduction schemes of Boson peaks…*

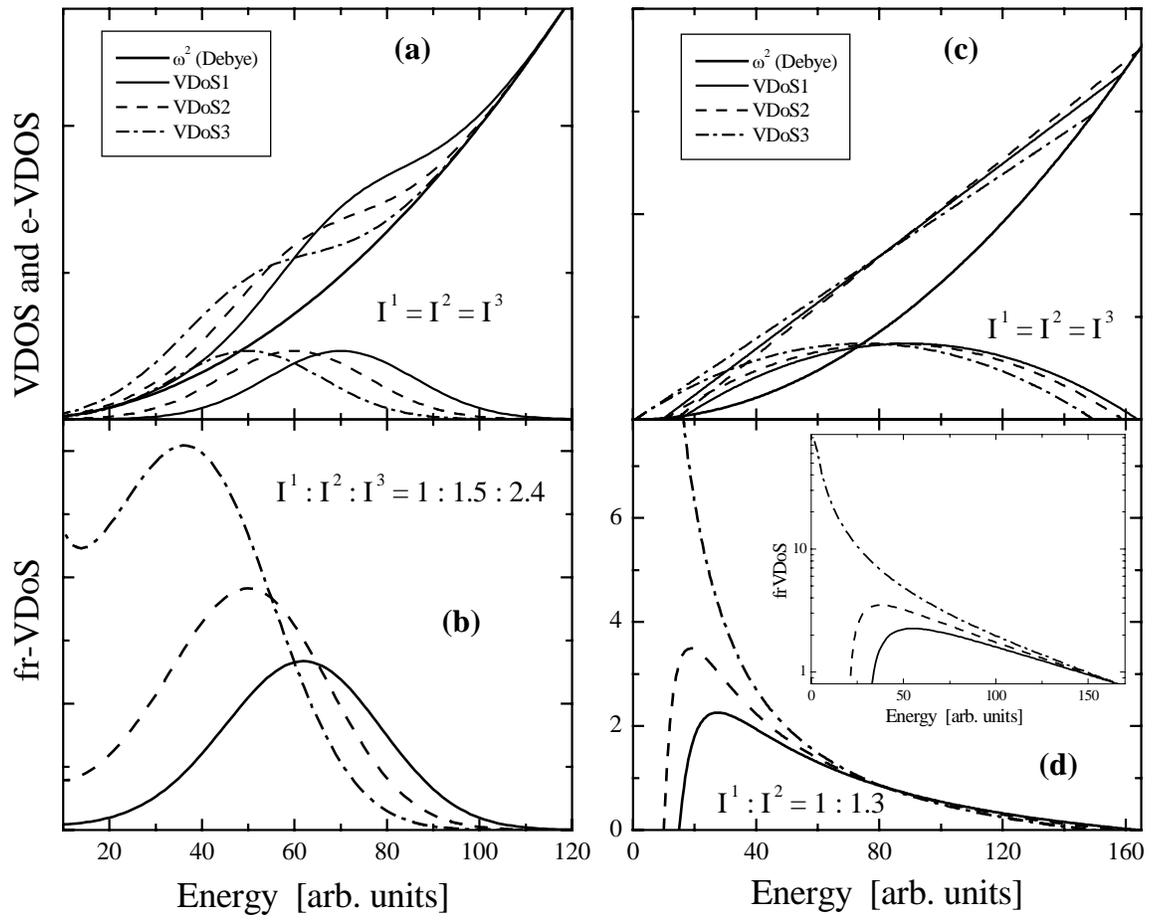

**Fig. 4**

*Reduction schemes of Boson peaks…*

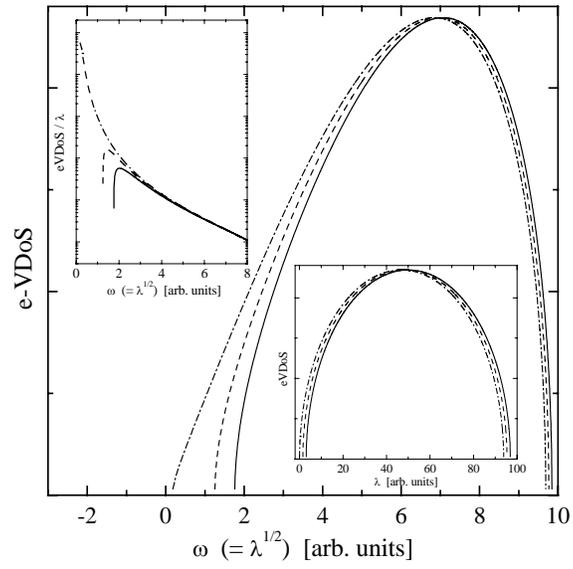

**Fig. 5**

*Reduction schemes of Boson peaks…*

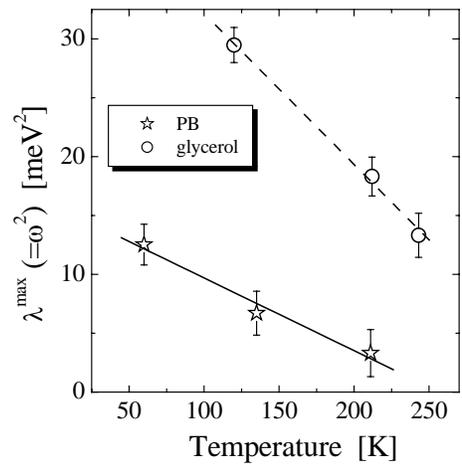

**Fig. 6**

*Reduction schemes of Boson peaks…*

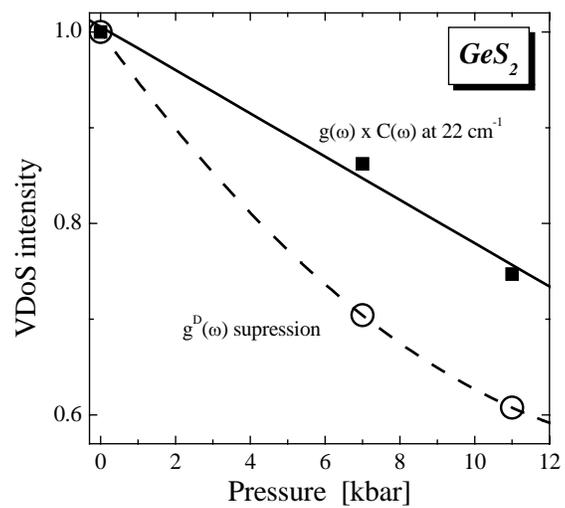

**Fig. 7**

*Reduction schemes of Boson peaks…*

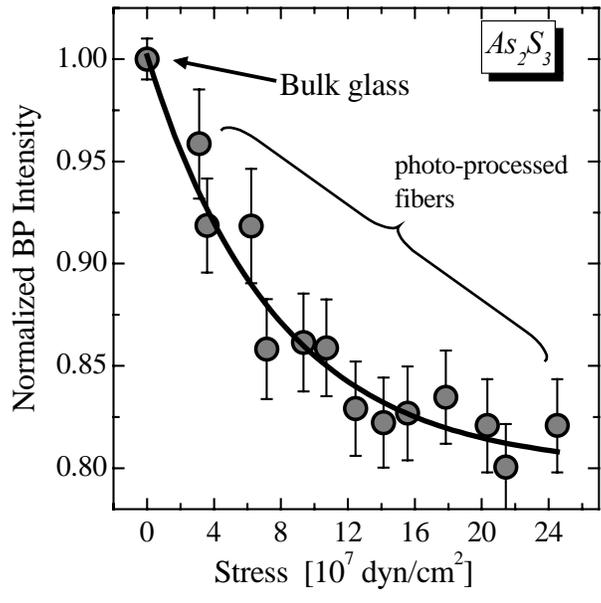

**Fig. 8**

*Reduction schemes of Boson peaks…*

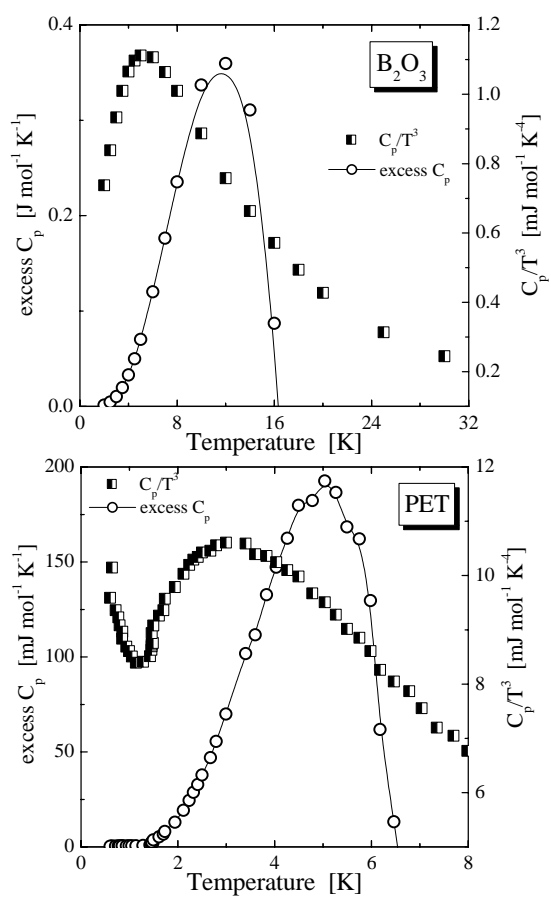

**Fig. 9**

*Reduction schemes of Boson peaks…*